\documentclass[11pt]{amsart}
\usepackage[utf8]{inputenc}
\usepackage[margin=1in]{geometry}
\usepackage{hyperref}
\usepackage{amsfonts}
\usepackage{amsmath}
\usepackage{amsthm}
\usepackage{amssymb}
\usepackage{mathrsfs}
\usepackage{tikz}
\usepackage[nameinlink,noabbrev]{cleveref}
\usepackage{amsaddr}
\usepackage{cleveref}

\DeclareFontFamily{OT1}{pzc}{}
\DeclareFontShape{OT1}{pzc}{m}{it}{<-> s * [1.200] pzcmi7t}{}
\DeclareMathAlphabet{\mathpzc}{OT1}{pzc}{m}{it}

\numberwithin{equation}{section}

\newcommand{\R}{\mathbb{R}}

\newcommand{\D}{\mathcal{D}}

\theoremstyle{plain}% Theorem-like structures provided by amsthm.sty

\theoremstyle{definition}

\theoremstyle{remark}

\title[]{The Fourier Uncertainty Principle Through the Lens \\ of Infinite Banded Matrices}
\author{Cameron L. Williams, Lara C. Ortelli, Gael Silva}

\begin{document}

\begin{abstract}
In this paper, we investigate the Fourier uncertainty principle through infinite banded matrices by way of a variational approach to minimizing the uncertainty product. Traditional methods for analyzing the Fourier uncertainty principle rely heavily on the Heisenberg--Weyl structure that underlie the Fourier transform and therefore do not generalize well beyond this setting. We provide three distinct approaches to developing the uncertainty principle based on the spectrum of a shifted quantum harmonic oscillator Hamiltonian, $H_{a,b}$. In the Hermite--Gauss basis, $H_{a,b}$ has an infinite, tridiagonal form which is central to the analysis. One approach exactly solves for the spectrum through analytic methods, another decomposes $H_{a,b}$ into an infinite sum of $2\times 2$ matrices, and the third approach is an algebraic approach to establishing the spectrum via a unitary equivalence with the quantum harmonic oscillator. The approaches taken herein provide alternate avenues that are more broadly applicable to other canonically conjugate operators for other integral transforms which maintain a tridiagonal form.
\end{abstract}

\maketitle

\section{Introduction} \label{sec:1}

The uncertainty principle is central to signal processing and quantum theory. The uncertainty principle constrains how well one can know the frequency content in a given time interval in a signal. There have been myriad treatments of the uncertainty principle \cite{Dubey,Folland_Uncertainty,Galindo_Pascual,Grochenig}, but most rely on the Lie algebraic properties of the non-commuting operators of position and momentum, specifically, that these operators commute to constant multiple of the identity. Folland and Sitaram's excellent survey of the uncertainty principle provides a variety of approaches to determining minimum uncertainty states \cite{Folland_Uncertainty}. Many of the traditional studies of the Fourier uncertainty principle rely on the group-theoretic properties of the Fourier transform to simplify the analysis significantly by transforming the uncertainty product into a zero-mean formulation. While not strictly necessary, this transformation does streamline the analysis.

This reliance on the algebraic structures inherent to the Fourier setting presents a challenge for more general uncertainty principles---even for other canonically conjugate operators for other unitary integral transforms---as many other pairs of non-commuting operators and integral transforms do not have such a nice structure. Analogues of uncertainty principles for other integral transforms have been explored \cite{Ghobber_Jaming,Hall,Rosler_Voit,Williams}. Some analogues mirror the usual Fourier uncertainty principle but assume a mean of zero to make the analysis tangible. Others are focused more on concentration of a function and its transform.

Developing techniques that apply more broadly and more faithfully reflect the Fourier uncertainty principle requires different techniques and approaches. This paper is devoted to developing some alternative methods that have broader applicability. Rather than relying on the Heisenberg--Weyl structure of the Fourier transform with exception to \Cref{sec:5}, our key ingredients will be the dilation property of the Fourier transform, that the position and momentum operators are canonically conjugate to each other via the Fourier transform, and the tridiagonal representation our operators have in the Hermite--Gauss basis.

In this paper, we present an alternative avenue to uncertainty minimization via a variational approach that leads to an eigenvalue equation for a shifted quantum harmonic oscillator Hamiltonian. Representing the local minimizers in the Hermite--Gauss basis leads to an infinite, tridiagonal banded matrix representation for the Hamiltonian. Thus, in a sense, the Hermite--Gauss basis is very close to the optimal orthonormal basis for analyzing the shifted quantum harmonic oscillator Hamiltonian. This alternative representation allows for the minimum uncertainty to be determined without relying on the specifics of the Heisenberg--Weyl algebra in favor of hard analysis and linear algebra methods. This approach can be applied to uncertainty principles for the canonically conjugate operators associated to other integral transforms.

The rest of the paper is organized as follows. In \Cref{sec:2}, we define the uncertainty product and then establish an eigenvalue equation for the local minimizers via a variational approach. In \Cref{sec:3}, we take an exact, analytic approach to solving for the entirety of the spectrum of our shifted Hamiltonian, including establishing its eigenfunctions. We then use this to develop an additive uncertainty principle which then gives the usual multiplicative uncertainty principle. In \Cref{sec:4}, we instead show that our Hamiltonian is bounded from below by $1$ via a decomposition into positive $2\times 2$ operators which in turn bounds the spectrum and gives the additive uncertainty principle. In \Cref{sec:5}, we develop a unitary operator that maps the usual quantum harmonic oscillator Hamiltonian to our shifted Hamiltonian which supplies precise knowledge of the spectrum of the shifted Hamiltonian and therefore the uncertainty principle. In \Cref{sec:6}, we discuss continuing work and further applications of the approach taken in \Cref{sec:4} to other integral transforms and non-commuting operators that will appear in upcoming work.

\section{The Uncertainty Product} \label{sec:2}

In this paper, we shall take the Fourier transform defined as an integral transform on $L^1(\R)\cap L^2(\R)$ by
\begin{equation}
\mathcal{F}f(y) = \int_{\R} \frac{1}{\sqrt{2\pi}} e^{-ixy} f(x)\,dx.
\end{equation}

\noindent The Fourier transform is an isometry $L^1(\R)\cap L^2(\R)$ into $L^2(\R)$ and can be extended abstractly to a unitary on $L^2(\R)$ in the standard way. For sake of simplicity of argument, we will work with the $L^2(\R)$ completion of the Fourier transform rather than be too caught up in details of the domain of definition.

In lieu of relying on the translation and modulation operators connected to the Heisenberg--Weyl structure of the Fourier transform, we will make significant use its dilation structure instead. To this end, let $\alpha > 0$. We define the dilation operator, denoted by $\mathcal{D}_{\alpha}$ on $L^2(\R)$ by
\begin{equation}
\mathcal{D}_{\alpha} f(x) = \sqrt{\alpha} f(\alpha x).
\end{equation}

\noindent $\D_{\alpha}$ is unitary on $L^2(\R)$. The Fourier transform on $L^2(\R)$ and the dilation operator play nicely together as seen in the following identity:
\begin{equation}
\mathcal{F} \mathcal{D}_{\alpha} = \mathcal{D}_{\alpha^{-1}} \mathcal{F}.
\end{equation}

\noindent Unlike the usual translation and modulation operators whose behavior is effectively unique to the Fourier transform as a consequence of the Stone--von Neumann theorem \cite[Ch. 14]{Hall}, this identity is due only to the homogeneity of the Lebesgue measure and the fact that the Fourier kernel is a function of $xy$, allowing for dilations to be transferred between the two variables.

In conjunction with the dilation property, we will rely on the Hermite--Gauss functions to supply concrete representations for our operators. Let the Hermite--Gauss functions, $e_n$, be defined by
\begin{equation}
e_n(x) = \frac{(-1)^n}{\sqrt{2^n n!\sqrt{\pi}}} e^{x^2/2} \frac{d^n}{dx^n} e^{-x^2}.
\end{equation}

\noindent These are an orthonormal set of $L^2(\R)$ eigenfunctions for the Fourier transform.

We will make use of the standard multiplication/position operator $\mathcal{X}$ and differentiation/momentum operator $\mathcal{P}$ throughout. We will define $\mathcal{X}$ by $\mathcal{X}f(x) = xf(x)$ on those $f\in L^2(\R)$ for which $\|\mathcal{X}f\|$ is finite, and $\mathcal{P} = \mathcal{F}^{-1}\mathcal{X}\mathcal{F}$ defined accordingly. For nice enough functions $f$, e.g. Schwartz space functions, $\mathcal{P}f(x) = -if'(x)$.

Given $f\in L^2(\R)$ with $\|f\| = 1$, we can define expectation values with respect to it. Specifically, for any polynomial $g$, we can define the expectation values
\begin{align*}
\langle g(\mathcal{X})\rangle &= \int_{\R} g(x) |f(x)|^2\,dx, \\
\langle g(\mathcal{P})\rangle &= \int_{\R} g(y) |\mathcal{F}f(y)|^2\,dy.
\end{align*}

\noindent The usual notion of uncertainty (or variance) for a(n essentially) self-adjoint operator $A$ is encoded via $\langle (A-\langle A\rangle)^2\rangle = \langle A^2 \rangle - \langle A\rangle^2$. Even for the simple cases of $\mathcal{X}$ and $\mathcal{P}$, the expectation values of $\mathcal{X}$ and $\mathcal{P}$ may be infinite for $f\in L^2(\R)$, e.g.
\begin{equation*}
f(x) = \chi_{[-1,1]}(x) + \sqrt{\frac{2}{\pi}} \frac{\sin(x)}{x}.
\end{equation*}

\noindent $f$ is an eigenfunction of the Fourier transform, and a quick computation shows that $\langle \mathcal{X}\rangle$ is infinite, and likewise $\langle \mathcal{P}\rangle$ is also infinite. To have a sensible notion of uncertainty, we must have that $\langle X\rangle$ and $\langle P\rangle$ are both finite which constrains us to a smaller subspace of $L^2(\R)$.

We also need to consider those $f$ for which $\langle \mathcal{X}^2\rangle$ and $\langle \mathcal{P}^2\rangle$ are finite. Since normalized $f$ induce probability densities, a quick application of Jensen's inequality \cite[p. 109]{Folland_Real_Analysis} shows that we need not separately consider those $f$ for which $\langle \mathcal{X}\rangle$ and $\langle \mathcal{X}^2\rangle$ are finite. Rather, we can assume that only $\langle \mathcal{X}^2\rangle$ is finite since $\varphi(x) = x^2$ is convex:
\begin{equation*}
\langle \mathcal{X}\rangle^2 \le \langle \mathcal{X}^2\rangle < \infty.
\end{equation*}

\noindent To this end, we define the subspace $\mu_2(\R)$ of those $f\in L^2(\R)$ for which $\langle \mathcal{X}^2\rangle$, and \emph{a fortiori} $\langle \mathcal{X}\rangle$, is finite:
\begin{equation}
\mu_2(\R) = \bigg\{ f\in L^2(\R) : \int_{\R} x^2 |f(x)|^2\,dx < \infty \bigg\}.
\end{equation}

\noindent This is not a closed subspace of $L^2(\R)$ which can be seen by simple example, however it is a closed subspace when endowed with the norm $\displaystyle \|f\|_{\mu_2}^2 = \int_{\R} (1+x^2)|f(x)|^2\,dx$. $\mu_2(\R)$ is also a dense subspace of $L^2(\R)$ in the usual norm as it contains the Schwartz space.

The Fourier uncertainty principle for $f\in \mu_2(\R)$ with $\mathcal{F}f\in \mu_2(\R)$ and $\|f\| = 1$ is often presented as
\begin{equation*}
\bigg(\int_{\R} (x-\langle \mathcal{X}\rangle)^2 |f(x)|^2\,dx\bigg) \bigg(\int_{\R} (y-\langle \mathcal{P}\rangle)^2 |\mathcal{F}f(y)|^2\,dy\bigg) \ge \frac{1}{4},
\end{equation*}

\noindent which can be rewritten as
\begin{equation}
\langle (\mathcal{X}-\langle \mathcal{X}\rangle)^2\rangle \langle (\mathcal{P}-\langle \mathcal{P}\rangle)^2\rangle \ge \frac{1}{4}
\end{equation}

\noindent which is the product of the variances in the original and Fourier domains. Notably, a uniform lower bound exists independent of $f$, meaning that regardless of choice of $f$, the uncertainty product cannot dip below $\frac{1}{4}$. Other uncertainty principles have explicit dependence on $f$ for their lower bounds, c.f. the Robertson--Schr\"{o}dinger inequality \cite[Ch. 12]{Hall}.

For the sake of future simplicity, we shall consider the following generalization of the uncertainty product
\begin{equation}
\bigg(\int_{\R} (x-a)^2 |f(x)|^2\,dx\bigg) \bigg(\int_{\R} (y-b)^2 |\mathcal{F}f(y)|^2\,dy\bigg) = \langle (\mathcal{X}-a)^2\rangle \langle (\mathcal{P}-b)^2\rangle,
\end{equation}

\noindent where $a,b\in\R$. When $a\neq \langle \mathcal{X}\rangle$, the expectation value $\langle (\mathcal{X}-a)^2\rangle$ exceeds $\langle (\mathcal{X}-\langle \mathcal{X}\rangle)^2\rangle_f$, and likewise for $\mathcal{P}$, so there is no harm in allowing for more generality as we will not risk introducing a smaller lower bound on the uncertainty product. Uncoupling $a$ and $b$ from $f$ will simplify the variational calculation significantly. While the resulting Hamiltonian-like equation would be effectively the same in either case, fixing $a$ and $b$ independently from $f$ from the outset reduces the complexity of the argument.

Let $a,b\in\R$ be fixed and $f,\mathcal{F}f\in \mu_2(\R)$. Define the uncertainty product functional $J$ by
\begin{equation}
J[f] = \bigg(\int_{\R} (x-a)^2 |f(x)|^2\,dx\bigg) \bigg(\int_{\R} (y-b)^2 |\mathcal{F}f(y)|^2\,dy\bigg).
\end{equation}

\noindent We wish to find a global minimum of this uncertainty product. To this end, we first look to the local extrema of $J$. Such a variational approach to uncertainty minimization is not new, c.f. \cite{Jackiw}, but our analysis diverges greatly from the standard approaches. Varying $J$ subject to the constraint $\|f\| = 1$ (or more generally $\|f\| < \infty$), its local extrema must satisfy
\begin{equation}
\sigma_\mathcal{P}^2 (x-a)^2 f + \sigma_\mathcal{X}^2 \mathcal{F}^{-1}((y-b)^2 \mathcal{F}f) = \lambda f,
\end{equation}

\noindent where $\sigma_\mathcal{X} = \|(\mathcal{X}-a)f\|$ and $\sigma_\mathcal{P} = \|(\mathcal{P}-b)\mathcal{F}f\|$.

As is, the above equation represents a challenging integral equation to solve. However, using properties of the Fourier transform, we may rewrite it as
\begin{equation}
\sigma_\mathcal{P}^2 (\mathcal{X}-a)^2 f + \sigma_\mathcal{X}^2 (\mathcal{P}-b)^2 f = \lambda f.
\end{equation}

\noindent Letting $f = \D_{\alpha} g$ and noting that $\mathcal{X} \D_{\alpha} = \alpha^{-1} \D_{\alpha} \mathcal{X}$ and $\mathcal{P} \D_{\alpha} = \alpha \D_{\alpha} \mathcal{P}$, we obtain
\begin{equation*}
\alpha^{-2} \sigma_\mathcal{P}^2 (\mathcal{X}-\alpha a)^2 g + \alpha^2 \sigma_\mathcal{X}^2 (\mathcal{P} - \alpha^{-1} b)^2 g = \lambda g.
\end{equation*}

\noindent Dividing through by $\alpha^{-2} \sigma_\mathcal{P}^2$, picking $\alpha^2 = \frac{\sigma_\mathcal{P}}{\sigma_\mathcal{X}}$, and relabeling $a$, $b$, and $g$ leads to the simpler eigenvalue equation 
\begin{equation}
(\mathcal{X}-a)^2 f + (\mathcal{P}-b)^2 f = Ef.
\end{equation}

\noindent The dilation operator allowed us to simplify our eigenvalue equation into a more symmetric form. This symmetric form is critical for our future analysis. In its original asymmetric form, the representation of $\sigma_\mathcal{P}^2 (\mathcal{X}-a)^2 f + \sigma_\mathcal{X}^2 (\mathcal{P}-b)^2 f$ in the Hermite--Gauss basis would be pentadiagonal rather than tridiagonal, significantly complicating the analysis.

Define the shifted Hamiltonian operator $H_{a,b}$ for those $f\in L^2(\R)$ such that $f,\mathcal{F}f\in \mu_2(\R)$ by
\begin{equation}
H_{a,b}f = ((\mathcal{X}-a)^2 + (\mathcal{P}-b)^2)f.
\end{equation}

\noindent From our above analysis, the local minimizers of the uncertainty product are then eigenfunctions of the operator $H_{a,b}$. As such, the eigenvalue equation $H_{a,b} f = Ef$ is the primary equation we shall analyze. Taking an inner product with $f$ on both sides leads directly to an additive form of the uncertainty \cite[Ch. 2]{Grochenig}:
\begin{equation}
\langle H_{a,b}f,f\rangle = \langle (\mathcal{X}-a)^2 f,f\rangle + \langle (\mathcal{P}-b)^2 f,f\rangle = \sigma_{\mathcal{X}}^2 + \sigma_{\mathcal{P}}^2.
\end{equation}

When $a = 0 = b$, $H_{a,b}$ is exactly the quantum harmonic oscillator Hamiltonian, and the solutions to the eigenvalue equation $H_{0,0}f = Ef$  are well-known \cite[Ch. 11]{Hall}. In this case, the eigenvalues are $2n+1$ and the corresponding eigenfunctions are the Hermite--Gauss functions.

\section{An Analytic Approach for Determining the Spectrum of \texorpdfstring{$H_{a,b}$}{Ha,b}} \label{sec:3}

In the Hermite--Gauss basis, the operators $\mathcal{X}$ and $\mathcal{P}$ have very simple forms:
\begin{align}
\mathcal{X} e_n &= \sqrt{\frac{n}{2}} e_{n-1} + \sqrt{\frac{n+1}{2}} e_{n+1}, \label{eq:x_representation} \\
\mathcal{P} e_n &= i\sqrt{\frac{n}{2}} e_{n-1} - i\sqrt{\frac{n+1}{2}} e_{n+1}. \label{eq:p_representation}
\end{align}

\noindent As $H_{a,b}$ is quadratic in $\mathcal{X}$ and $\mathcal{P}$, it will at most be pentadiagonal in the Hermite--Gauss basis. However, due to symmetry, a simple computation shows that it is in fact merely tridiagonal in the Hermite--Gauss basis and has the following representation:
\begin{equation}
H_{a,b} = \begin{pmatrix} \alpha^2 + 1 & 2\sqrt{\frac{1}{2}}\alpha & & & & \\ 2\sqrt{\frac{1}{2}} \alpha & \alpha^2 + 3 & 2\sqrt{1} \alpha & & & \\ & 2\sqrt{1} \alpha & \alpha^2 + 5 & 2\sqrt{\frac{3}{2}} \alpha & & \\ & & 2\sqrt{\frac{3}{2}} \alpha & \alpha^2 + 7 & 2\sqrt{2} \alpha & \\ & & \qquad\ddots & \qquad\ddots & \qquad\ddots \end{pmatrix}.
\end{equation}

\noindent Solving the eigenvalue equation $H_{a,b} x = Ex$ leads to solving the infinite tridiagonal system
\begin{equation}
\begin{pmatrix} \alpha^2 + 1 - E & 2\sqrt{\frac{1}{2}}\alpha & & & & \\ 2\sqrt{\frac{1}{2}} \alpha & \alpha^2 + 3 - E & 2\sqrt{1} \alpha & & & \\ & 2\sqrt{1} \alpha & \alpha^2 + 5 - E & 2\sqrt{\frac{3}{2}} \alpha & & \\ & & 2\sqrt{\frac{3}{2}} \alpha & \alpha^2 + 7 - E & 2\sqrt{2} \alpha & \\ & & \qquad\ddots & \qquad\ddots & \qquad\ddots \end{pmatrix} \begin{pmatrix} x_0 \\ \vspace*{-0.5em} \\ x_1 \\ \vspace*{-0.5em} \\ x_2 \\ \vspace*{-0.5em} \\ x_3 \\ \vdots \end{pmatrix} = \begin{pmatrix} 0 \\ \vspace*{-0.5em} \\ 0 \\ \vspace*{-0.5em} \\ 0 \\ \vspace*{-0.5em} \\ 0 \\ \vdots \end{pmatrix}
\end{equation}

\noindent It is clear that $x_0$ will be a free parameter upon a quick inspection of the system. $x_0$ can then be constrained by requiring that the vector $x$ be $L^2$ normalized to $1$.

We have a three-term recurrence relation for the coefficients:
\begin{equation}
2\sqrt{\frac{n}{2}} \alpha x_{n-1} + (\alpha^2 + 2n+1-E) x_n + 2 \sqrt{\frac{n+1}{2}} \alpha x_{n+1} = 0.
\end{equation}

\noindent Here we take $x_{-1} = 0$ and $x_0$ free so that if $n = 0$, this reduces to a two-term recurrence relation as expected. Determining the first few values of $x_n$ in terms of $x_0$ makes the following explicit form for the $x_n$ apparent which can be verified with a slightly tedious induction argument
\begin{equation}
x_n = \frac{(-1)^n x_0}{\sqrt{2^n n!} \alpha^n} \sum_{m=0}^n 2^m \binom{n}{m} \alpha^{2(n-m)} \prod_{l=0}^{m-1} \bigg(l + \frac{1}{2} - \frac{E}{2}\bigg).
\end{equation}

\noindent Using the definition of the gamma function, this can be simplified to
\begin{equation}
x_n = \frac{x_0}{\sqrt{2^n n!} \alpha^n} \sum_{m=0}^n 2^m \binom{n}{m} \alpha^{2(n-m)} \frac{\Gamma\big(m + \frac{1}{2} - \frac{E}{2}\big)}{\Gamma\big(\frac{1}{2} - \frac{E}{2}\big)}.
\end{equation}

When $E = 1, 3, 5, \ldots$, the product will eventually terminate and thus $x_n$ will have a maximum number of summands. Specifically, if $E = 2k+1$, $x_n$ will have a maximum of $k$ summands. This of course does not guarantee that $x$ converges in $\ell^2$, but it does lead to a simpler convergence analysis. A simplified expression for $x_n$ exists when $E = 2k+1$:
\begin{equation}
x_n = \frac{(-1)^n x_0}{\sqrt{2^n n!}\alpha^n} \sum_{m=0}^n (-1)^m 2^m \binom{n}{m} \alpha^{2(n-m)} \frac{k!}{(k-m)!} = \frac{(-1)^{k+n} 2^k k! x_0}{\sqrt{2^n n!}} \alpha^{n-2k} L_k^{(n-k)}\bigg(\frac{\alpha^2}{2}\bigg),
\end{equation}

\noindent where $L_k^{(n)}$ is the generalized Laguerre polynomial \cite[p. 1000]{Gradshteyn_Ryzhik}. When $k=0$, $x_n$ matches exactly the basis coefficients for the harmonic oscillator coherent states. Furthermore, from the identity \cite{Erdelyi}
\begin{equation*}
\sum_{n=0}^{\infty} \frac{1}{n!} t^n L_k^{(n-k)}(t)^2 = \frac{1}{k!} t^k e^t,
\end{equation*}

\noindent we have that the $x_n$ are square summable and sum to
\begin{equation*}
\sum_{n=0}^{\infty} |x_n|^2 = \sum_{n=0}^{\infty}  \bigg(\frac{(-1)^{k+n} 2^k k! x_0}{\sqrt{2^n n!}} \alpha^{n-2k} L_k^{(n-k)}\bigg(\frac{\alpha^2}{2}\bigg)\bigg)^2 = \frac{2^k k! x_0^2}{\alpha^{2k}} e^{\alpha^2/2}.
\end{equation*}

\noindent Normalizing $x$ to $1$ gives
\begin{equation}
x = (-1)^{k+n} \sqrt{2^k k!} e^{-\alpha^2/4} \sum_{n=0}^{\infty} \frac{1}{\sqrt{2^n n!}} \alpha^{n-k} L_k^{(n-k)}\bigg(\frac{\alpha^2}{2}\bigg) e_n.
\end{equation}

\noindent Note that $x_n$ is a polynomial in $\alpha$, i.e. it has no negative powers of $\alpha$.

On the other hand, if $E \neq 1, 3, 5, \ldots$, the number of summands for $x_n$ trails off to infinity. As such, $x_n$ itself could diverge and prevent $x$ from being in $\ell^2$ at all and indeed this is the case. Let $x_{n,m}$ represent the $m$\textsuperscript{th} summand for $x_n$. The $x_{n,m}$ have the same sign for all $m > -E + \frac{1}{2}$. Furthermore, for fixed $\alpha > 0$ and $E > 0$,
\begin{equation}
\bigg|\frac{x_{n,m+1}}{x_{n,m}}\bigg| = \alpha^{-2} \frac{n-m}{m+1} |2m + 1 - E|.
\end{equation}

\noindent For large $n$, this is greater than $1$ for most $m$ between $1$ and $n$ so $x_{n,m}$ is mostly increasing as a function of $m$. Additionally, from Stirling's approximation, $|x_{n,\lfloor n/2\rfloor}| \sim 2^{n/2}$. As the $x_{n,m}$ are eventually the same sign, increasing in $m$, and $|x_{n,\lfloor n/2\rfloor}| \sim 2^{n/2}$, upon summing, $x_n$ is superexponential in $n$ and so $x$ is not in $\ell^2$. Thus, for any $E\neq 1, 3, 5, \ldots$, the eigenvector $x$ is not in $\ell^2$.

The above analysis shows that the eigenvalue problem $H_{a,b}x = Ex$ has $\ell^2$ solutions if and only if $E = 1, 3, 5, \ldots$. It remains to see that for any fixed $\alpha$ these eigenvectors are in fact a basis for $\ell^2$. We do this by showing that any vector that is orthogonal to all of these vectors must be the zero vector. Suppose $y$ is orthogonal to every aforementioned eigenvector of $H_{a,b}$, then $y = (y_0, y_1, y_2, \ldots)$ solves the system
\begin{equation*}
(-1)^{k+n} \sqrt{2^k k!} e^{-\alpha^2/4} \sum_{n=0}^{\infty} \frac{y_n}{\sqrt{2^n n!}} \alpha^{n-k} L_k^{(n-k)}\bigg(\frac{\alpha^2}{2}\bigg) = 0
\end{equation*}

\noindent for all $k\ge 0$, or equivalently
\begin{equation*}
\sum_{n=0}^{\infty} \widetilde{y}_n L_k^{(n-k)}(\beta) = 0
\end{equation*}

\noindent for all $k\ge 0$, $\widetilde{y}_n$ which is a scalar multiple of $y_n$, and fixed $\beta\ge 0$.

The Laguerre polynomials $L_k^{(n-k)}(x)$ are all of degree $k$ and the first $k+1$ are linearly independent. All others after the first $k+1$ are linearly dependent. If we define $z_k^{(n)} = L_k^{(n-k)}(\beta)$, $y$ solves the infinite system
\begin{equation}
\begin{pmatrix} z_1^{(1)} & z_2^{(1)} & z_3^{(1)} & \cdots \\ z_1^{(2)} & z_2^{(2)} & z_3^{(2)} & \cdots  \\ z_1^{(3)} & z_2^{(3)} & z_3^{(3)} & \cdots \\ \vdots & \vdots & \vdots & \ddots \end{pmatrix} \begin{pmatrix} \widetilde{y}_0 \\ \widetilde{y}_1 \\ \widetilde{y}_2 \\ \vdots \end{pmatrix} = \begin{pmatrix} 0 \\ 0 \\  0 \\ \vdots \end{pmatrix}
\end{equation}

\noindent However, since $L_k^{(n-k)}$ is linearly dependent on the first $k+1$ Laguerre polynomials when $n\ge k+1$, we can simplify this to a triangular form via Gaussian elimination along the columns to obtain a lower triangular form
\begin{equation*}
\begin{pmatrix} z_1^{(1)} & 0 & 0 & \cdots \\ z_1^{(2)} & \widetilde{z}_2^{(2)} & 0 & \cdots  \\ z_1^{(3)} & \widetilde{z}_2^{(3)} & \widetilde{z}_3^{(3)} & \cdots \\ \vdots & \vdots & \vdots & \ddots \end{pmatrix} \begin{pmatrix} \widetilde{y}_0 \\ \widetilde{y}_1 \\ \widetilde{y}_2 \\ \vdots \end{pmatrix} = \begin{pmatrix} 0 \\ 0 \\  0 \\ \vdots \end{pmatrix}
\end{equation*}

\noindent With this triangular form, it is clear that $\widetilde{y}_n = 0$ for all $n$ and so $y_n = 0$ for all $n$. Thus, the eigenvectors of $H_{a,b}$ form an orthonormal basis for $\ell^2$. Even though $\alpha$ (and therefore $\beta$) is fixed, the linear dependence still carries over in the usual way.

This then gives the additive uncertainty principle nearly immediately. Let $x\in \ell^2$, then $x$ can be represented as $x = \sum_n c_n x_n$. The additive uncertainty is then expectation value $\langle x, H_{a,b}x\rangle$ which can be established directly
\begin{equation}
\langle x, H_{a,b}x\rangle = \sigma_\mathcal{X}^2 + \sigma_\mathcal{P}^2 = \sum_{n=0}^{\infty} (2n+1) |c_n|^2 \ge \sum_{n=0}^{\infty} 1\cdot |c_n|^2 = \langle x,x\rangle = 1. \label{eq:minimum_additive_uncertainty}
\end{equation}

\noindent Thus $\sigma_\mathcal{X}^2 + \sigma_\mathcal{P}^2 \ge 1$.

Employing the dilation operator as in \cite{Folland_Uncertainty}, we can obtain the multiplicative uncertainty principle from the additive uncertainty principle. Since $\sigma_\mathcal{X}^2 + \sigma_\mathcal{P}^2 \ge 1$ for any suitable $f$, dilating $f$ with $\D_{\alpha}$ leads to
\begin{equation*}
\alpha^2 \sigma_\mathcal{X}^2 + \alpha^{-2} \sigma_\mathcal{P}^2 \ge 1.
\end{equation*}

\noindent Optimizing the expression on the left as a function of $\alpha$ to find an $\alpha$ that gives the smallest additive uncertainty for suitable $f$ leads again to $\alpha^2 = \frac{\sigma_\mathcal{P}}{\sigma_\mathcal{X}}$. Making this substitution in the additive uncertainty principle, we have the standard multiplicative uncertainty
\begin{equation}
\sigma_\mathcal{X}^2 \sigma_\mathcal{P}^2 \ge \frac{1}{4}.
\end{equation}

\section{A \texorpdfstring{$2\times 2$}{2 x 2} Matrix Decomposition Approach for Bounding the Spectrum of \texorpdfstring{$H_{a,b}$}{Ha,b}} \label{sec:4}

In the previous section, we established the precise spectral values of $H_{a,b}$, but in general this can be a very difficult task, even for tridiagonal operators. Any self-adjoint operator on finite-dimensional vector spaces can be converted to tridiagonal form \cite[Ch. 8.3]{Golub_Van_Loan}, so if it were possible to determine the spectra of tridiagonal operators in general, it would be possible to determine the spectra of any self-adjoint operator.

We can instead find a uniform lower bound for $H_{a,b}$, i.e. we show the existence of a $\delta > 0$ independent of $\alpha$ such that $H_{a,b} - \delta I \ge 0$. This does not require that we know the precise spectrum of $H_{a,b}$ but does constrain its spectrum. To do this, we decompose $H_{a,b} - \delta I$ as an infinite sum of positive (zero-padded) $2\times 2$ matrices. If $H_{a,b} - \delta I \ge 0$, then by evaluating $\langle e_0, (H_{a,b} - \delta I) e_0\rangle$, we know that the upper left entry of $H_{a,b} - \delta I$ must be nonnegative, i.e. $\alpha^2 + 1 - \delta \ge 0$ for all $\alpha \ge 0$. Since we want a uniform $\delta$ that is independent of $\alpha$, we see immediately that $\delta \le 1$.

In fact, $\delta = 1$ is optimal. $H_{a,b} - I$ has the form
\begin{equation}
H_{a,b} - I = \begin{pmatrix} \alpha^2 & 2\sqrt{\frac{1}{2}}\alpha & & & & \\ 2\sqrt{\frac{1}{2}} \alpha & \alpha^2 + 2 & 2\sqrt{1} \alpha & & & \\ & 2\sqrt{1} \alpha & \alpha^2 + 4 & 2\sqrt{\frac{3}{2}} \alpha & & \\ & & 2\sqrt{\frac{3}{2}} \alpha & \alpha^2 + 6 & 2\sqrt{2} \alpha & \\ & & \qquad\ddots & \qquad\ddots & \qquad\ddots \end{pmatrix}.
\end{equation}

\noindent We define a sequence of $2\times 2$ matrices $M_k$ for $k\ge 1$ by
\begin{equation}
M_k = \begin{pmatrix} \alpha^2 & 2\sqrt{\frac{k}{2}} \alpha \\ 2\sqrt{\frac{k}{2}} \alpha & 2k \end{pmatrix}.
\end{equation}

\noindent Each $M_k$ is a positive operator by virtue of being self-adjoint and Sylvester's criterion as its upper left entry is positive and its determinant is $0$. Alternatively, $M_k$ is self-adjoint and has eigenvalues $0$ and $\alpha^2 + 2k$.

Let $\widetilde{M}_k$ be an operator on $\ell^2$ defined by
\begin{equation}
(\widetilde{M}_k)_{i,j} = \begin{cases} 0, & i, j \not\in \{k,k+1\} \\ (M_k)_{i-k,j-k}, & i, j \in \{k,k+1\} \end{cases},
\end{equation}

\noindent i.e. $\widetilde{M}_k$ is constructed by placing $M_k$ in the $(k,k)$ to $(k+1,k+1)$ position and zero padding appropriately. $\widetilde{M}_k$ is still positive as $M_k$ is positive and lies along the block diagonal. Moreover, $H_{a,b} - I$ can be reconstructed from $\widetilde{M}_k$ by summing, i.e.
\begin{equation}
H_{a,b} - I = \sum_{k=1}^{\infty} \widetilde{M}_k.
\end{equation}

\noindent Here we take the convergence in the weak sense on $c_{00}\subseteq \ell^2$, meaning
\begin{equation*}
\langle x, (H_{a,b} - I)x\rangle = \bigg \langle x, \bigg(\sum_{k=1}^{\infty} \widetilde{M}_k\bigg) x\bigg\rangle = \sum_{k=1}^{\infty} \langle x, \widetilde{M}_k x\rangle
\end{equation*}

\noindent for all finitely supported $x$. This can be extended further to a larger set of $x$ for which the basis coefficients are not eventually zero but rather decay sufficiently rapidly, e.g. faster than $k^{-1}$.

Since $\displaystyle \sum_{k=1}^{\infty} \widetilde{M}_k$ is a positive operator, so too is $H_{a,b} - I$, and thus the spectrum of $H_{a,b} - I$ must be nonnegative. Furthermore, we can conclude that the spectrum of $H_{a,b}$ is contained in $[1,\infty)$ in agreement with the results in \Cref{sec:3}. Since $H_{a,b} - I$ is positive, we have that $\langle x, H_{a,b} x\rangle \ge \langle x, x\rangle$ as in \eqref{eq:minimum_additive_uncertainty}. While this method does not give the precise spectral values, it does provide a streamlined argument for a lower bound on $H_{a,b}$ and thus a lower bound for the additive, and therefore multiplicative, uncertainty.

\section{An Algebraic Approach for Determining the Spectrum of \texorpdfstring{$H_{a,b}$}{Ha,b}} \label{sec:5}

As $H_{a,b}$ and the usual quantum harmonic oscillator, $H_{0,0}$, are isospectral, it is natural to search for a unitary operator, $\mathcal{U}_{a,b}$, that transforms one into the other. It is easiest to see how this unitary is defined by establishing it a piece at a time. We wish to map $\mathcal{X}^2$ to $(\mathcal{X}-a)^2$ and similarly $\mathcal{P}^2$ to $(\mathcal{P}-b)^2$ by conjugating with $\mathcal{U}_{a,b}$. Given the appearance of squares throughout, it is reasonable to require the simpler conditions that $\mathcal{U}_{a,b}$ maps $\mathcal{X}$ to $\mathcal{X}-a$ and $\mathcal{P}$ to $\mathcal{P}-b$. We will first work with $\mathcal{X}$ then with $\mathcal{P}$ separately for the sake of simplicity. To this end, we will search for \emph{two} unitaries, $\mathcal{U}_a$ and $\mathcal{U}_b$, whose composition will be the unitary $\mathcal{U}_{a,b}$. The defining relations are then $\mathcal{U}_a \mathcal{X} \mathcal{U}_a^* = \mathcal{X}-a$ and $\mathcal{U}_b \mathcal{P} \mathcal{U}_b^* = \mathcal{P}-b$, and we further require that $\mathcal{U}_a \mathcal{P} \mathcal{U}_a^* = \mathcal{P}$ and $\mathcal{U}_b \mathcal{X} \mathcal{U}_b^* = \mathcal{X}$.

To simplify the search for $\mathcal{U}_a$, we assume that $\mathcal{U}_a$ is generated by some (essentially) self-adjoint operator $\mathcal{S}$ that is independent of $a$. We shall assume that the domains and ranges of $\mathcal{X}$ and $\mathcal{S}$ are sufficiently dense in each other so that the formal operations below are sensible. The defining property $\mathcal{U}_a \mathcal{X} \mathcal{U}_a^* = \mathcal{X}-a$ becomes
\begin{equation}
e^{ia\mathcal{S}} \mathcal{X} e^{-ia\mathcal{S}} = \mathcal{X}-a.
\end{equation}

\noindent However, from Baker--Campbell--Hausdorff, we obtain the relation
\begin{equation}
\mathcal{X} + ia[\mathcal{S},\mathcal{X}] - \frac{a^2}{2!} [\mathcal{S},[\mathcal{S},\mathcal{X}]] + \ldots = \mathcal{X}-a.
\end{equation}

\noindent As $S$ is independent of $a$, we must have that the higher order commutators are all $0$, i.e. $\mathcal{S}$ commutes with $[\mathcal{S},\mathcal{X}]$. Moreover, $i[\mathcal{S},\mathcal{X}] = -1$ so that $[\mathcal{S},\mathcal{X}] = i$ which guarantees that the higher order commutators are $0$. This does not uniquely define $\mathcal{S}$ as any polynomial in $\mathcal{X}$ could be added to $\mathcal{S}$ and still retain the commutator identity. Furthermore, from the requirement $\mathcal{U}_a \mathcal{P} \mathcal{U}_a^* = \mathcal{P}$, we have
\begin{equation}
\mathcal{P} + ia [\mathcal{S},\mathcal{P}] - \frac{a^2}{2!} [\mathcal{S},[\mathcal{S},\mathcal{P}]] + \cdots = \mathcal{P}.
\end{equation}

\noindent We conclude that $\mathcal{S}$ and $\mathcal{P}$ commute.

Similarly, if we define $\mathcal{U}_b = \exp(ib\mathcal{T})$ with the domains and ranges of $\mathcal{T}$ and $\mathcal{P}$ suitably dense in each other, we have that $\mathcal{T}$ must commute with $[\mathcal{T},\mathcal{P}]$ and $[\mathcal{T},\mathcal{P}] = -i$. $\mathcal{T}$ is again not uniquely defined by these commutator identities. Likewise, the condition $\mathcal{U}_b \mathcal{X} \mathcal{U}_b^* = \mathcal{X}$ forces $\mathcal{T}$ and $\mathcal{X}$ to commute.

A simple way to ensure that $[\mathcal{S},\mathcal{X}] = i$, $\mathcal{S}$ and $\mathcal{P}$ commute, $[\mathcal{T},\mathcal{P}] = -i$, and $\mathcal{T}$ and $\mathcal{X}$ commute is to let $\mathcal{S} = -\mathcal{P}$ and $\mathcal{T} = \mathcal{X}$. The Stone--von Neumann theorem in fact states that these are effectively the only operators satisfying the above identities.

Conjugating the harmonic oscillator Hamiltonian $H_{0,0}$ with $\mathcal{U}_a$ leads to
\begin{equation}
\mathcal{U}_a (\mathcal{X}^2 + \mathcal{P}^2) \mathcal{U}_a^* = (\mathcal{X}-a)^2 + \mathcal{P}^2.
\end{equation}

\noindent Conjugating with $\mathcal{U}_b$ then gives
\begin{equation}
\mathcal{U}_b ((\mathcal{X}-a)^2 + \mathcal{P}^2) \mathcal{U}_b^* = (\mathcal{X}-a)^2 + (\mathcal{P}-b)^2.
\end{equation}

\noindent Conjugating the Hamiltonian $H_{0,0}$ with $\mathcal{U}_b \mathcal{U}_a$ gives exactly the Hamiltonian $H_{a,b}$ as desired which in turn justifies the observation that $H_{0,0}$ and $H_{a,b}$ have the same spectrum.

The operator $\mathcal{U} = \mathcal{U}_b \mathcal{U}_a = e^{ib\mathcal{X}} e^{-ia\mathcal{P}}$ is a well-known operator. Through Baker--Campbell--Hausdorff, we can express this as a single, simple exponential as $\mathcal{X}$ and $\mathcal{P}$ commute to a constant multiple of the identity:
\begin{equation}
e^{ib\mathcal{X}} e^{-ia\mathcal{P}} = e^{ib\mathcal{X} - ia\mathcal{P} - i\frac{ab}{2}}.
\end{equation}

\noindent Since $-\frac{1}{2}iab$ is constant, we can effectively ignore it as an overall phase factor. Returning to our original definition of $\alpha$ as $a+ib$, $a = \frac{1}{2}(\alpha + \overline{\alpha})$ and $b = \frac{1}{2i}(\alpha - \overline{\alpha})$ and the right side becomes
\begin{equation}
e^{\frac{1}{2}(\alpha(\mathcal{X}-i\mathcal{P}) - \overline{\alpha}(\mathcal{X}+i\mathcal{P}))}.
\end{equation}

This is exactly the displacement operator in the topic of coherent states, typically denoted $D(\alpha)$, \cite[p. 10]{Klauder_Skagerstam} up to an overall scale factor. From this, we see that the displacement operator encodes the unitary equivalence of the Hamiltonians $H_{0,0}$ and $H_{a,b}$, and therefore the minimum additive uncertainty is given by the minimum eigenvalue of $H_{0,0}$ as noted previously, in accordance again with \eqref{eq:minimum_additive_uncertainty}. The displacement operator also provides a simple closed-form expression for the eigenfunctions of $H_{a,b}$: they are merely given by $D(\alpha) e_n$.

Through this unitary equivalence, it is easy to see that ladder operators exist for $H_{a,b}$ as well and generalize the usual quantum harmonic oscillator ladder operators. The ladder operators for $H_{a,b}$ are $\mathcal{L}_{a,b} = \mathcal{U} \mathcal{L}_{0,0} \mathcal{U}^*$ and $\mathcal{R}_{a,b} = \mathcal{U} \mathcal{R}_{0,0} \mathcal{U}^*$, where $\mathcal{L}_{0,0}$ and $\mathcal{R}_{0,0}$ are the usual quantum harmonic oscillator lowering and raising operators, respectively:
\begin{align}
\mathcal{L}_{a,b} &= (\mathcal{X}-a) + i(\mathcal{P}-b), \\
\mathcal{R}_{a,b} &= (\mathcal{X}-a) - i(\mathcal{P}-b).
\end{align}

\noindent $\mathcal{L}_{a,b}$ and $\mathcal{R}_{a,b}$ are (formal) adjoints of each other. We then have decompositions of $H_{a,b}$ in these ladder operators:
\begin{align}
H_{a,b} &= \mathcal{R}_{a,b} \mathcal{L}_{a,b} + 1, \\
&= \mathcal{L}_{a,b} \mathcal{R}_{a,b} - 1.
\end{align}

\noindent The existence of these ladder operators makes it simple to establish the spectrum of $H_{a,b}$, even independent of the knowledge of the existence of the unitary equivalence with $H_{0,0}$.

Unlike the previous two sections, the methods of this section are uniquely applicable to the Fourier setting. Inherent to the above analysis is the Heisenberg--Weyl structure of the Fourier transform. The unitary equivalence of the two Hamiltonians being encoded by the displacement operator is a defining feature of the quantum harmonic oscillator and the Fourier uncertainty principle.

\section{Continuing Work} \label{sec:6}

The methods in \Cref{sec:4} are more broadly applicable than in just the Fourier transform case. The first author has applied this to analyzing the nonzero mean uncertainty principle for a generalization of the Fourier--Bessel transform to the whole line analogous to \cite{Rosler_Voit} which will appear in an upcoming paper. Further work has been done to extend the methods of \Cref{sec:4} to more general pairs of non-commuting operators that play nicely with dilations with modified representations from \eqref{eq:x_representation} and \eqref{eq:p_representation} and will appear in a separate upcoming paper.

While the algebraic approach of \Cref{sec:5} are unique to the Fourier transform and quantum harmonic oscillator, the analytic approach in \Cref{sec:3} may be more broadly applicable to other pairs of non-commuting operators. Even in the Fourier--Bessel case, this approach is tricky and not obvious. An investigation into which pairs of non-commuting operators this method may be useful for would be an interesting and useful endeavor.

\bibliographystyle{plain}
\bibliography{references}

\end{document}